\documentclass[letterpaper,twocolumn,american,showpacs,pre,aps,superscriptaddress]{revtex4}
\usepackage[latin1]{inputenc}
\usepackage{bm}
\usepackage{multirow,amssymb,amsbsy,amsmath}
\usepackage{stmaryrd}
\usepackage{graphicx}
\usepackage{hyperref}
\makeatletter
\usepackage{pifont}
\makeatother

\begin{document}

\title{Three-dimensional Monte Carlo simulations of the quantum linear
       Boltzmann equation}

\author{Heinz-Peter Breuer}

\email{breuer@physik.uni-freiburg.de}

\affiliation{Physikalisches Institut, Universit\"at Freiburg,
             Hermann-Herder-Strasse 3, D-79104 Freiburg, Germany}

\author{Bassano Vacchini}

\email{bassano.vacchini@mi.infn.it}

\affiliation{Dipartimento di Fisica
 dell'Universit\`a di Milano and INFN,
 Sezione di Milano,
 Via Celoria 16, 20133, Milan, Italy}

\date{\today}

\begin{abstract}
Recently the general form of a translation-covariant quantum
Boltzmann equation has been derived which describes the dynamics
of a tracer particle in a quantum gas. We develop a stochastic
wave function algorithm that enables full three-dimensional Monte
Carlo simulations of this equation. The simulation method is used
to study the approach to equilibrium for various scattering cross
sections and to determine dynamical deviations from Gaussian
statistics through an investigation of higher-order cumulants.
Moreover, we examine the loss of coherence of superpositions of
momentum eigenstates and determine the corresponding decoherence
time scales to quantify the transition from quantum to classical
behavior of the state of the test particle.
\end{abstract}



\maketitle

\section{Introduction}

In recent times major efforts have been devoted to the study and
understanding of the dynamics of open systems \cite{Breuer2007},
both in order to give a realistic, quantitative description of the
time evolution of a quantum system coupled to a generally larger
system considered as environment, as well as with the aim to
engineer suitable environments driving the dynamics of the quantum
system according to the will of the experimenter. When considering
an open system one can either use a phenomenological description
for the system-environment interaction, as well as for the
environment itself, or rely on a strictly microscopic description
of the physical system considered. While the first approach can be
more viable and flexible, the second is clearly of more
fundamental nature. In the present paper we will focus on the
second approach, performing a numerical study of a recently
observed \cite{Vacchini2000a,Hornberger2006b} quantum master
equation for the description of the motion of a quantum test
particle in a gas. Such a master equation is the quantum version
of the classical linear Boltzmann equation and gives a microscopic
description of the dynamics of the test particle, only relying on
the gas properties and on the detailed expression of the
interaction between test particle and gas particles. The quantum
linear Boltzmann equation is characterized by its covariance under
translations \cite{Vacchini2001b}, and has been used in a
simplified form for the quantitative description of experiments on
collisional decoherence
\cite{Hornberger2003a,Hornberger2004a,Vacchini2004a}. Quantitative
experiments, testing the transition from the quantum to the
classical world, are in fact typical situations in which a truly
microscopic description of the physical system of interest is
mandatory.

The translation-covariant quantum linear Boltzmann equation that
will be studied here may be written in the form of a quantum
master equation for the time-dependent density matrix $\rho(t)$ of
the test particle,
\begin{equation} \label{BOLTZMANN}
 \frac{d}{dt}\rho(t) = {\mathcal{L}}\rho(t).
\end{equation}
The infinitesimal generator ${\mathcal{L}}$ represents a
superoperator in Lindblad form \cite{Lindblad1976a,Gorini1976a}.
It is well known that such a quantum master equation with Lindblad
structure allows an unravelling through a stochastic process for
the state vector in the particle's Hilbert space. Here, we will
concentrate on the so-called quantum jump method in which the
state vector follows a piecewise deterministic process, consisting
of smooth, deterministic evolution periods and discontinuous,
random quantum jumps
\cite{Dalibard1992,Molmer1993a,Dum1992a,Carmichael1993a}. It will
be demonstrated that the application of this method to the quantum
Boltzmann equation (\ref{BOLTZMANN}) is indeed feasible and leads
to a simple and numerically efficient three-dimensional Monte
Carlo simulation technique for the dynamical behavior of the
particle.

By means of this technique we will study in particular relaxation
properties of this master equation for various microscopic
scattering cross sections, as well as deviations from Gaussian
statistics. Given the quantum nature of the equation we will also
investigate the time evolution of quantum superposition states.

The paper is organized as follows. In Sec.~\ref{QME} we briefly
introduce the quantum linear Boltzmann equation together with its
basic features, while in Sec.~\ref{MONTE-CARLO} we show how to
apply the Monte Carlo wave function method to this master
equation, which has a very intuitive physical meaning when working
in the momentum representation. In Sec.~\ref{RESULTS} we describe
the numerical algorithms used and present the simulation results.
Besides studying relaxation dynamics and deviation from Gaussian
statistics, we will focus on decoherence effects in momentum
space, comparing relaxation and decoherence rates and providing
analytical estimates for the latter. Finally, in
Sec.~\ref{sec:conclusions} we comment on our results and point out
possible extensions and generalizations of the work.

\section{Quantum master equation}\label{QME}
We briefly introduce the quantum master equation that we want to
study numerically, together with its most relevant features,
referring the reader to
\cite{Vacchini2000a,Vacchini2001a,Vacchini2001b,Petruccione2005a,Hornberger2006b}
for a more detailed presentation. The explicit form of the
Lindblad generator of Eq.~(\ref{BOLTZMANN}) is given by
\begin{multline} \label{1}
 {\mathcal{L}}\rho(t) = -\frac{i}{\hbar} [H_0,\rho(t)]  +
 \frac{n_{\mathrm{gas}}}{m_*^2}\int d\mathbf{Q}\, \sigma (\mathbf{Q})
\\
\times
 \Big[ e^{\frac{i}{\hbar}\mathbf{Q}\cdot{\mathbf{X}}}\sqrt{S
      (\mathbf{Q},{\mathbf{P}})}\rho(t) \sqrt{S
      (\mathbf{Q},{\mathbf{P}})}e^{-\frac{i}{\hbar}\mathbf{Q}\cdot\mathbf{X}}
\\
 - \frac{1}{2} \left\{ {S (\mathbf{Q},{\mathbf{P}})},\rho(t) \right\} \Big],
\end{multline}
where $H_0={\mathbf{P}}^2/2M$ is the kinetic energy of the test
particle with mass $M$, ${\mathbf{X}}$ and ${\mathbf{P}}$ are its
position and momentum operators respectively, $n_{\mathrm{gas}}$
is the number density of the gas, $m$ is the mass of the gas
particles, and $m_*=mM/ (m+M)$ denotes the reduced mass.

The master equation (\ref{BOLTZMANN}) with the generator (\ref{1}) is
the quantum version of the classical linear Boltzmann equation,
provided collisions can be described in Born approximation, or the
scattering cross section only depends on the momentum transfer
experienced by the test particle in the collision. We denote such a
scattering cross section by $\sigma (\mathbf{Q})$. A more general
expression of the quantum linear Boltzmann equation, including the
full scattering cross section on general grounds, has been obtained in
\cite{Hornberger2006b}, and relies on the appearance of the
operator-valued scattering amplitude. It coincides with
Eq.~\eqref{BOLTZMANN} in Born approximation or when the full
scattering cross section only depends on the momentum transfer. Such a
master equation describes the motion of a quantum test particle
interacting through collisions with a dilute gas of environmental
particles.  The positive quantity ${S(\mathbf{Q},{\mathbf{P}})}$, here
appearing operator-valued, is a two-point correlation function of the
gas related to the spectrum of its density fluctuations. It is most
often expressed in its dependence on the momentum transfer
$\mathbf{Q}$ and on the energy transfer
$E(\mathbf{Q},\mathbf{P})=Q^2/2M+\mathbf{P}\cdot\mathbf{Q}/M$
characterizing a collision in which the test particle gains a momentum
$\mathbf{Q}$, changing its momentum from $\mathbf{P}$ to
$\mathbf{P}+\mathbf{Q}$. For the case of a free gas of particles
described by a Maxwell-Boltzmann distribution it is explicitly given
by
\begin{align}
   \label{eq:mb}
   S (\mathbf{Q},\mathbf{P})
&=S (\mathbf{Q},E (\mathbf{Q},\mathbf{P}))
\\
\nonumber &= \sqrt{\frac{\beta m}{2\pi}}\frac{1}{Q}\exp \left[
-\frac{\beta}{8m}\frac{(2mE
(\mathbf{Q},\mathbf{P})+Q^2)^2}{Q^2}\right],
\end{align}
where $\beta=1/k_{\mathrm{B}}T$ is the inverse temperature of the
gas.

In full generality this two-point correlation function known as
dynamic structure factor \cite{Schwabl2003,Pitaevskii2003} is
given by the Fourier transform with respect to momentum transfer
$\mathbf{Q}$ and energy transfer $E(\mathbf{Q},\mathbf{P})$ of the
time dependent density-density autocorrelation function of the
medium,
\begin{equation}
   \label{eq:2}
   S (\mathbf{Q},E)=\frac{1}{2\pi\hbar}\int dt \int d
   \mathbf{X}\,
  e^ {\frac{i}{\hbar}(E t -
        \mathbf{Q}\cdot\mathbf{X})}
G (\mathbf{X},t),
\end{equation}
where
\begin{equation}
   \label{eq:1}
    G (\mathbf{X},t)=\frac{1}{N}\int d\mathbf{Y} \,
        \left \langle
         N(\mathbf{Y})
         N(\mathbf{X}+\mathbf{Y},t)
         \right \rangle
\end{equation}
describes density correlations in the gas. While for the general
case of an interacting system of particles an exact evaluation is
obviously not feasible, the dynamic structure factor has several
important properties that are helpful in the construction of a
phenomenological ansatz. An important property of the dynamic
structure factor granting the existence of the expected canonical
stationary solution of \eqref{1} is the so-called detailed balance
condition according to which
\begin{equation}
   \label{eq:3}
    S (\mathbf{Q},E)=e^{-\beta E} S (-\mathbf{Q},-E).
\end{equation}

Another crucial feature of Eq.~\eqref{1} is its covariance under
translations. Considering the unitary representation
$U(\mathbf{a})=\exp({-{i}\mathbf{a}\cdot{{\mathbf{P}}}}/\hbar)$,
$\mathbf{a}\in \mathbb{R}^3$, of the group of three-dimensional
space translations in the test particle's Hilbert space, one has
that
\begin{equation}
   \label{eq:33}
   \mathcal{L}[U
(\mathbf{a})\rho(t) U^{\scriptscriptstyle \dagger} (\mathbf{a})]=U
(\mathbf{a})\mathcal{L}[\rho(t)]U^{\scriptscriptstyle \dagger}
(\mathbf{a}).
\end{equation}
In fact, Eq.~\eqref{1} complies with the general structure of
translation-covariant master equations obtained by Holevo
\cite{Holevo1996a}, providing a physically relevant example of
this general mathematical structure. The property of covariance
reflects the underlying symmetry under translations, arising
because we are considering a homogeneous gas and the interaction
potential between test particle and gas particles only depends on
the relative distance between the two. An important consequence of
this property, that we shall exploit in the simulations carried
out in Sec.~\ref{RESULTS}, is the fact that the algebra generated
by the momentum operators ${\mathbf{P}}$ is invariant under the
time evolution described by $\mathcal{L}$.

The master equation \eqref{1} is an operator equation, coinciding
with the classical linear Boltzmann equation as far as the
diagonal matrix elements in the momentum representation are
concerned, but also describing quantum coherences corresponding to
the off-diagonal matrix elements, as well as the time evolution of
highly non classical motional states, such as superposition
states. The Lindblad structure of such a quantum linear Boltzmann
equation apart from preservation of trace and positivity is of
great importance in that it implies the possibility to consider
suitable stochastic unravellings leading to efficient Monte Carlo
simulations of the time evolution of different quantities of
physical interest, despite the high complexity of the problem.

\section{Monte Carlo wave function method}\label{MONTE-CARLO}

We give a short description of the standard stochastic jump
unravelling
\cite{Dalibard1992,Molmer1993a,Dum1992a,Carmichael1993a} of the
master equation (\ref{BOLTZMANN}). For further details on the
theory and the numerical implementation see Ref.~\cite{Breuer2007}
and references therein. Introducing the Lindblad operators
\begin{equation}
 L(\mathbf{Q}) = e^{\frac{i}{\hbar}\mathbf{Q}\cdot\mathbf{X}}
 \sqrt{\frac{n_{\mathrm{gas}}}{m_*^2}\sigma (\mathbf{Q}) S(\mathbf{Q},\mathbf{P})}
\end{equation}
we can write the Boltzmann equation as follows,
\begin{multline}\label{QMEQ}
 \frac{d}{dt}\rho (t) = -\frac{i}{\hbar} [H_0,\rho (t)] \\
 + \int d {\mathbf{Q}} \left[ L(\mathbf{Q}) \rho (t) L^{\dagger}(\mathbf{Q})
 - \frac{1}{2} \left\{ L^{\dagger}(\mathbf{Q})L(\mathbf{Q}),\rho (t) \right\} \right].
\end{multline}
This equation leads to a stochastic unravelling through a
piecewise deterministic process in Hilbert space. This means
that the realizations $|\psi(t)\rangle$ of the process consist of
deterministic evolution periods (deterministic drift) which are
interrupted by discontinuous changes of the state vector (quantum
jumps).

\subsection{Simulation algorithm}\label{sec:simulation-algorithm}
The realizations of the piecewise deterministic process are
defined through the following algorithm. In between two jumps the
state vector follows a deterministic time-evolution which is given
by the nonlinear Schr\"odinger equation
\begin{equation}
 \frac{d}{dt}|\psi\rangle = \left[ -\frac{i}{\hbar} H_{\mathrm{eff}}
 + \frac{1}{2} \int d\mathbf{Q}\, \langle \psi | L^{\dagger}(\mathbf{Q})L(\mathbf{Q}) | \psi \rangle
 \right] |\psi\rangle
\end{equation}
with the non-Hermitian Hamiltonian
\begin{equation}
\label{eff}
 H_{\mathrm{eff}} = H_0 - \frac{i\hbar}{2} \int d\mathbf{Q}\, L^{\dagger}(\mathbf{Q})L(\mathbf{Q}).
\end{equation}
Suppose that at time $t_0$ a jump into some state
$|\psi(t_0)\rangle \equiv |\psi\rangle$ occurred. The total rate
for jumps out of this state is given by
\begin{equation} \label{GAMMA-TOTAL}
 \Gamma(|\psi\rangle) = \int d\mathbf{Q}\,
 \langle \psi | L^{\dagger}(\mathbf{Q})L(\mathbf{Q}) | \psi \rangle.
\end{equation}
Hence, the next jump will take place at time $t_0+\tau$, where
$\tau$ is a stochastic time step which follows the cumulative
distribution function
\begin{equation} \label{WAITING-TIME-DISTR}
 F(\tau) = 1 - || \exp[ -iH_{\mathrm{eff}}\tau/\hbar ] |\psi\rangle ||^2.
\end{equation}
This is the waiting time distribution, i.~e., $F(\tau)$ represents
the probability that the next jump takes place somewhere in the
interval $(t_0,t_0+\tau)$. Employing the inversion method, for
instance, one determines the stochastic time step $\tau$ by
solving the equation
\begin{equation}
 || \exp[ -iH_{\mathrm{eff}}\tau/\hbar ] |\psi\rangle ||^2 = \eta
\end{equation}
for $\tau$, where $\eta$ is a random number uniformly distributed
over the interval $(0,1)$.

Once the random time step has been determined one carries out a
jump of the state vector $|\psi(t_0+\tau)\rangle \equiv
|\psi\rangle$ at time $t_0+\tau$ by the replacement
\begin{equation}
\label{jump}
 |\psi\rangle \longrightarrow
 \frac{L(\mathbf{Q})|\psi\rangle}{||L(\mathbf{Q})|\psi\rangle||}.
\end{equation}
The momentum transfer $\mathbf{Q}$ is to be drawn from the
probability density
\begin{equation}
 R(\mathbf{Q}) = \frac{\langle
\psi|L^{\dagger}(\mathbf{Q})L(\mathbf{Q})|\psi\rangle}{\Gamma(|\psi\rangle)},
\end{equation}
which is normalized as
\begin{equation}
 \int d\mathbf{Q}\, R(\mathbf{Q}) = 1.
\end{equation}

The process thus defined represents a stochastic unravelling of
the quantum master equation in the sense that the expectation
value
\begin{equation}
 \rho(t) = {\mathbb{E}}\big[ |\psi(t)\rangle\langle\psi(t)| \big]
\end{equation}
yields a solution of Eq.~(\ref{QMEQ}). In the Monte Carlo wave
function method one numerically generates large samples of
realizations and estimates all desired quantities with the help of
appropriate sample averages.

\subsection{Momentum representation}\label{sec:moment-repr}
As already mentioned in Sec.~\ref{QME} an important property of
Eq.~\eqref{QMEQ} is its covariance under the action of the
translation group. This implies that the algebra generated by the
three commuting momentum operators ${\mathbf{P}}$, i.~e. the
generators of translations, is left invariant under the action of
the master equation. A function of the momentum operators goes
over with elapsing time to another function of the momentum
operators only, when evolving according to the master
equation~\eqref{QMEQ}. The considered unravelling of
Eq.~\eqref{QMEQ} preserves this property in the sense that linear
combinations of improper eigenvectors $|\mathbf{P}\rangle$ of the
three commuting momentum operators are preserved in form in each
single realization, which is of great advantage in the
simulations. In fact $H_{\mathrm{eff}}$ given by~\eqref{eff} is
only a function of the momentum operators, thus simply acting in a
multiplicative way on the momentum eigenvectors, while the jumps
effected by the Lindblad operators $L(\mathbf{Q})$ according
to~\eqref{jump} simply correspond to shifts
\begin{displaymath}
   |\mathbf{P}\rangle \rightarrow |\mathbf{P}+\mathbf{Q}\rangle.
\end{displaymath}

Given a master equation covariant under an Abelian symmetry group
it is generally true that the algebra generated by the commuting
self-adjoint operators which act as generators of the symmetry is
left invariant under time evolution. Correspondingly one can
consider unravellings leaving invariant in form linear
combinations of common eigenvectors of the generators of the
symmetry, where the jumps only lead to a shift between different
eigenvectors. For an initial state given by such an eigenvector
the stochastic unravelling leads to a pure jump process. Consider
for example the master equation for the damped harmonic oscillator
with Lindblad operators $a$ and $a^{\dagger}$, covariant under the
group $U (1)$
\cite{Holevo1993a,Holevo1995b,Vacchini2002b,Vacchini2007c}, where
the generator of the symmetry is the number operator
$N=a^{\dagger}a$. In this case one can consider a stochastic
unravelling given by a suitable piecewise deterministic process,
where the jumps effected by the Lindblad operators $a$ and
$a^{\dagger}$ in the single realizations are simply given by the
shifts $|n\rangle\rightarrow |n-1\rangle$ and
$|n\rangle\rightarrow |n+1\rangle$, respectively
\cite{Breuer2007}.

For the case of a generic initial state the algorithm used to
generate a realization of the process may conveniently be
expressed in term of the wave function
$\tilde{\psi}(\mathbf{P})=\langle\mathbf{P}|\psi\rangle$ in the
momentum representation. In fact, for a state $|\psi(t_0)\rangle
\equiv |\psi\rangle$ the total transition rate given by
Eq.~(\ref{GAMMA-TOTAL}) takes the form
\begin{equation} \label{eq:13}
 \Gamma(|\psi\rangle) = \int d\mathbf{P} \, \Gamma
 (\mathbf{P})|\tilde{\psi} (\mathbf{P})|^2,
\end{equation}
where
\begin{equation} \label{eq:4}
 \Gamma({\mathbf{P}})=\frac{n_{\mathrm{gas}}}{m_*^2}\int d\mathbf{Q}\,
 \sigma (\mathbf{Q}){S (\mathbf{Q},{\mathbf{P}})}
\end{equation}
is the total rate for transitions out of a state characterized by
the momentum $\mathbf{P}$. The deterministic time evolution in
between the quantum jumps determined by the effective Hamiltonian
\eqref{eff} is explicitly given by
\begin{equation} \label{eq:12}
   |\psi(t_0+\tau)\rangle = \frac{\int d\mathbf{P} \,
   e^{-iH_0 (\mathbf{P})\tau/\hbar}e^{-\Gamma (\mathbf{P})\tau/2}\tilde{\psi}
   (\mathbf{P})|\mathbf{P}\rangle}{\sqrt{\int d\mathbf{P} \,
   e^{-\Gamma (\mathbf{P})\tau}|\tilde{\psi} (\mathbf{P})|^2}},
\end{equation}
while the waiting time distribution defined by
Eq.~(\ref{WAITING-TIME-DISTR}) becomes
\begin{equation} \label{eq:14}
 F(\tau) = 1 - \int d\mathbf{P} \, e^{-\Gamma(\mathbf{P})\tau}|\tilde{\psi} (\mathbf{P})|^2.
\end{equation}
The jumps are described by
\begin{equation}
   \label{eq:15}
    |\psi\rangle \longrightarrow \frac{\int d\mathbf{P} \,
    \sqrt{S(\mathbf{Q},\mathbf{P})}\tilde{\psi}(\mathbf{P})
    |\mathbf{P}+\mathbf{Q}\rangle}{\sqrt{\int d\mathbf{P}
    \,{S (\mathbf{Q},\mathbf{P})} |\tilde{\psi} (\mathbf{P})|^2}},
\end{equation}
where the momentum transfers $\mathbf{Q}$ follow the distribution
\begin{equation}
   \label{eq:16}
 R(\mathbf{Q}) = \frac{n_{\mathrm{gas}}}{m^2_{\ast}}
 \frac{\int d\mathbf{P} \, \sigma (\mathbf{Q}) S(\mathbf{Q},\mathbf{P})
 |\tilde{\psi}(\mathbf{P})|^2}{\int d\mathbf{P} \, \Gamma (\mathbf{P})|\tilde{\psi} (\mathbf{P})|^2}.
\end{equation}
As it immediately appears from Eqs.~\eqref{eq:12}-\eqref{eq:16}
the choice of an initial state given by a finite linear
superposition of improper momentum eigenvectors leads to very
important simplifications (see Sec.~\ref{sec:decoherence}).

\subsection{Determination of the total transition rate}\label{sec:trans-rate}
The total transition rate (\ref{eq:4}) plays an important role in
the simulation algorithm. It can be analytically calculated in
several interesting cases, e.~g. considering a gas of free
particles described by Maxwell-Boltzmann statistics so that
\begin{multline}
   \label{eq:6}
   \Gamma(\mathbf{P})=\frac{n_{\mathrm{gas}}}{m_*^2} \sqrt{\frac{\beta
       m}{2\pi}} \int d\mathbf{Q}\, \sigma (\mathbf{Q})
\\
\times \frac{1}{Q}\exp \left[ -\frac{\beta}{8m} \frac{(2mE
(\mathbf{Q},\mathbf{P})+Q^2)^2}{Q^2}\right].
\end{multline}
It is of great advantage to introduce the scaled momenta
\begin{equation}
   \label{eq:7}
   \mathbf{K}=\frac{\mathbf{Q}}{m_* v_{\mathrm{mp}}}
\end{equation}
and
\begin{equation}
   \label{eq:8}
   \mathbf{U}=\frac{\mathbf{P}}{M v_{\mathrm{mp}}},
\end{equation}
where $v_{\mathrm{mp}}=\sqrt{2/m\beta}$ is the most probable
velocity of the gas particles. In terms of these quantities the
function $\Gamma$, now expressed in terms of the scaled momentum
$\mathbf{U}$, becomes
\begin{multline}
   \label{eq:99}
   \Gamma (U) =n_{\mathrm{gas}}v_{\mathrm{mp}}2\sqrt{\pi}
   \int_0^{\infty} dK\, K\,\sigma (K)
\\
\times \int_{-1}^{+1} d\xi\, \exp\left[
      -\left(\frac{K}{2}+U\xi\right)^2 \right],
\end{multline}
where the variable $\xi$ denotes the cosine between the vectors
$\mathbf{U}$ and $\mathbf{K}$, and the scattering cross section
$\sigma(K)$ has been supposed to depend only on the
modulus of the momentum transfer. Integrating over $\xi$ one has
\begin{multline}
   \label{eq:9}
   \Gamma (U) = n_{\mathrm{gas}}v_{\mathrm{mp}}\frac{\pi}{U}
   \int_0^{\infty} dK\, K\,\sigma (K)
\\
\times\left[\mathrm{erf}\left
      (\frac{K}{2}+U\right) -\mathrm{erf}\left (\frac{K}{2}-U\right)
\right] .
\end{multline}
\begin{figure}[htb]
\begin{center}
\includegraphics[width=0.8\linewidth]{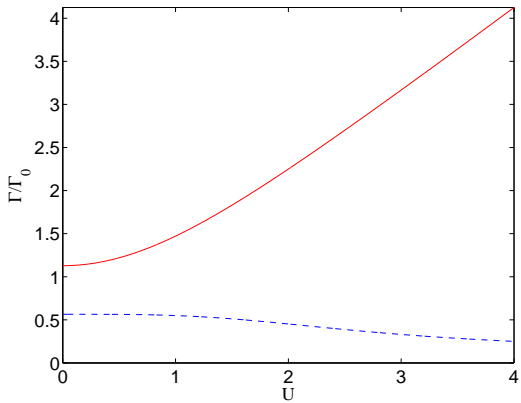}
\caption{(Color online) The total transition rate $\Gamma(U)$ for
a constant cross section [Eq.~(\ref{GAMMA-CONSTANT})] (continuous
line) and for a Gaussian cross section with $a=1$
[Eq.~(\ref{GAMMA-GAUSS})] (broken line).\label{transition-rate}}
\end{center}
\end{figure}
For the case of a constant scattering cross section,
\begin{equation}
 \sigma(K) = \sigma = \mbox{const},
\end{equation}
the total transition rate is found to be
\begin{equation} \label{GAMMA-CONSTANT}
 \Gamma(U) = \Gamma_0 \left\{ \left[1+2U^2\right] \frac{{\mathrm{erf}}(U)}{2U}
 + \frac{1}{\sqrt{\pi}}e^{-U^2}\right\},
\end{equation}
where the quantity
\begin{equation}
 \Gamma_0 = n_{\mathrm{gas}} v_{\mathrm{mp}} 4\pi \sigma
\end{equation}
represents the total scattering rate corresponding to an incoming
flux of particles with the most probable velocity
$v_{\mathrm{mp}}$. For a Gaussian scattering cross section of the
form
\begin{equation}
 \sigma(K) = \sigma\, e^{-aK^2/4}
\end{equation}
one has instead
\begin{equation} \label{GAMMA-GAUSS}
 \Gamma(U) = \frac{\Gamma_0}{aU}
 \left\{ {\mathrm{erf}}(U) - \frac{\mathrm{erf}(U/\sqrt{a+1})}{\sqrt{a+1}}
 e^{-\frac{a}{a+1}U^2}
 \right\}.
\end{equation}

The form of these functions is illustrated in
Fig.~\ref{transition-rate}. One observes that for large $U$ the
function $\Gamma(U)$ increases linearly with $U$ in the case of a
constant cross section, while it decreases as $U^{-1}$ in the case
of a Gaussian cross section. We note also that in terms of the
scaled momentum variables the canonical mean value
$\langle{\mathbf{P}}^2/2M\rangle_{\mathrm{eq}}={3}/2\beta$ of the
kinetic energy of the test particle reached at thermal equilibrium
corresponds to $\langle{\mathbf{U}}^2\rangle_{\mathrm{eq}}=3m/2M$.

\section{Numerical algorithms and simulation results}\label{RESULTS}

\subsection{Dissipative effects}
We first study relaxation to thermal equilibrium by considering
the dynamics of an ensemble of momentum eigenstates of a test
particle in a Maxwell-Boltzmann gas. Hence, using the scaled
momentum variables introduced in Sec.~\ref{QME} we investigate
initial states of the form
\begin{equation}
 |\psi(0)\rangle = |{\mathbf{U}}(0)\rangle.
\end{equation}
The application of the quantum jump unravelling described in
Sec.~\ref{MONTE-CARLO} then leads to a classical stochastic
process ${\mathbf{U}}(t)$ for the test particle momentum. This
process is a pure jump process the realizations of which are
obtained through the algorithm described below. Note that this
algorithm corresponds to the standard algorithm that is used for
the stochastic simulation of classical Markovian master equations
\cite{Gillespie1992}. In order to study relaxation to a Gaussian
thermal state we will consider the behavior in time of first and
second moments of the momentum distribution, as well as of various
cumulants of the distribution.

\subsubsection{Simulation method}
The state ${\mathbf{U}}(0)$ at the initial time $t=0$ is to be
drawn from a given distribution for the initial data. Suppose a
jump occurred at some time $t_0$ leading to the state
${\mathbf{U}}(t_0)\equiv {\mathbf{U}}$. The next jump will then
take place at time $t_0+\tau$, where $\tau$ is a stochastic time
step which is given by
\begin{equation} \label{DEFTAU}
 \tau = -\frac{1}{\Gamma(U)} \ln \eta.
\end{equation}
$\eta$ is a random number which is uniformly distributed over the
interval $(0,1)$, and $\Gamma(U)$ represents the total transition
rate. Since the process is a pure jump process, ${\mathbf{U}}$
stays constant between $t_0$ and $t_0+\tau$.

At time $t_0+\tau$ one carries out a jump by replacing
\begin{equation}
 {\mathbf{U}} \longrightarrow {\mathbf{U}} + \frac{m_*}{M}
 {\mathbf{K}}.
\end{equation}
The momentum transfer ${\mathbf{K}}$ is determined as follows.
First, one draws random numbers $(K,\xi)$ that follow the joint
probability density $R(K,\xi)$ which is normalized as
\begin{equation}
 \int_0^{\infty} dK  \int_{-1}^{+1} d\xi\, R(K,\xi) = 1.
\end{equation}
$K$ is the size of the momentum transfer and $\xi$ the cosine of
the angle between ${\mathbf{K}}$ and ${\mathbf{U}}$. In the Monte
Carlo simulations shown below we have used the rejection method to
determine $(K,\xi)$. Second, one draws a uniformly distributed
random unit vector ${\mathbf{e}}$. Then, the momentum transfer is
given by the formula
\begin{equation}
 {\mathbf{K}} = {\mathbf{K}}_{\parallel} + {\mathbf{K}}_{\perp}
 = K\xi\frac{\mathbf{U}}{U} +
 K\sqrt{1-\xi^2}\frac{{\mathbf{U}}\times{\mathbf{e}}}{|{\mathbf{U}}\times{\mathbf{e}}|}.
\end{equation}
${\mathbf{K}}_{\parallel}$ is the component of ${\mathbf{K}}$
which is parallel to the particle momentum ${\mathbf{U}}$, and
${\mathbf{K}}_{\perp}$ its component perpendicular to it.
Repeating these steps until the desired final time $t_f$ is
reached one obtains a realization of the process ${\mathbf{U}}(t)$
over the whole time interval $[0,t_f]$.

\subsubsection{Constant scattering cross section}
\begin{figure}[htb]
\begin{center}
\includegraphics[width=0.8\linewidth]{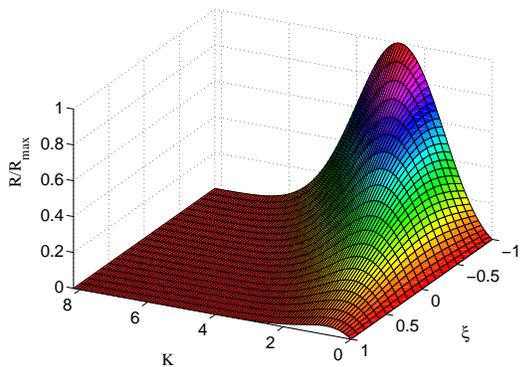}
\caption{(Color online) The probability density $R(K,\xi)$ given
by Eq.~(\ref{R-K-XI}) for $U=1$.\label{jump-prob}}
\end{center}
\end{figure}

We first address momentum and energy relaxation for the case of a
constant scattering cross section $\sigma$. The corresponding
total transition rate $\Gamma(U)$ is given by
Eq.~(\ref{GAMMA-CONSTANT}), while the joint probability of
$(K,\xi)$ takes the form
\begin{equation} \label{R-K-XI}
 R(K,\xi) = \frac{\Gamma_0}{2\sqrt{\pi}\Gamma(U)}
 K \exp\left[ -\left(\frac{K}{2}+U\xi\right)^2 \right].
\end{equation}
This probability density is illustrated in Fig.~\ref{jump-prob}.
For small $U$ the density $R(K,\xi)$ is nearly uniform in $\xi$,
corresponding to an isotropic distribution, while it has a
pronounced maximum at $\xi=-1$ if $U$ is not small. In the latter
case there is thus a strong tendency that the momentum transfer
${\mathbf{K}}$ is opposite to the direction of the particle
momentum ${\mathbf{U}}$.

\begin{figure}[htb]
\begin{center}
\includegraphics[width=0.8\linewidth]{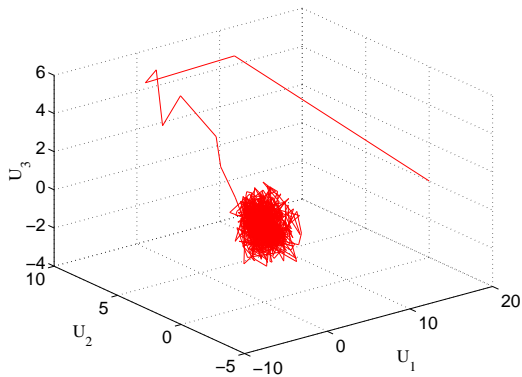}
\caption{(Color online) A single realization of the process
${\mathbf{U}}(t)$ for $m/M = 1$.\label{fig-real}}
\end{center}
\end{figure}

\begin{figure}[htb]
\begin{center}
\includegraphics[width=0.8\linewidth]{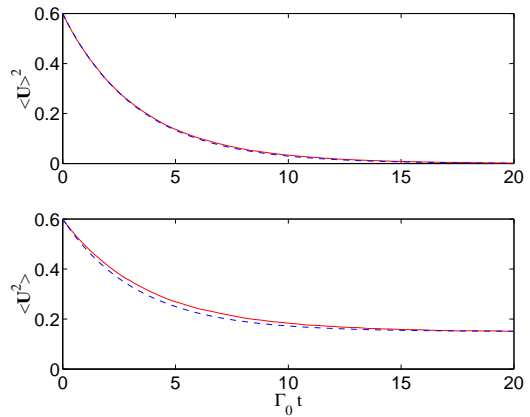}
\caption{(Color online) Averages over $10^4$ realizations for a
constant cross section and $m/M=0.1$. The broken lines represent
the approximate relaxation dynamics according to Eqs.
(\ref{APPROX-1}) and (\ref{APPROX-2}), respectively.
\label{ensemble-1}}
\end{center}
\end{figure}

\begin{figure}[htb]
\begin{center}
\includegraphics[width=0.8\linewidth]{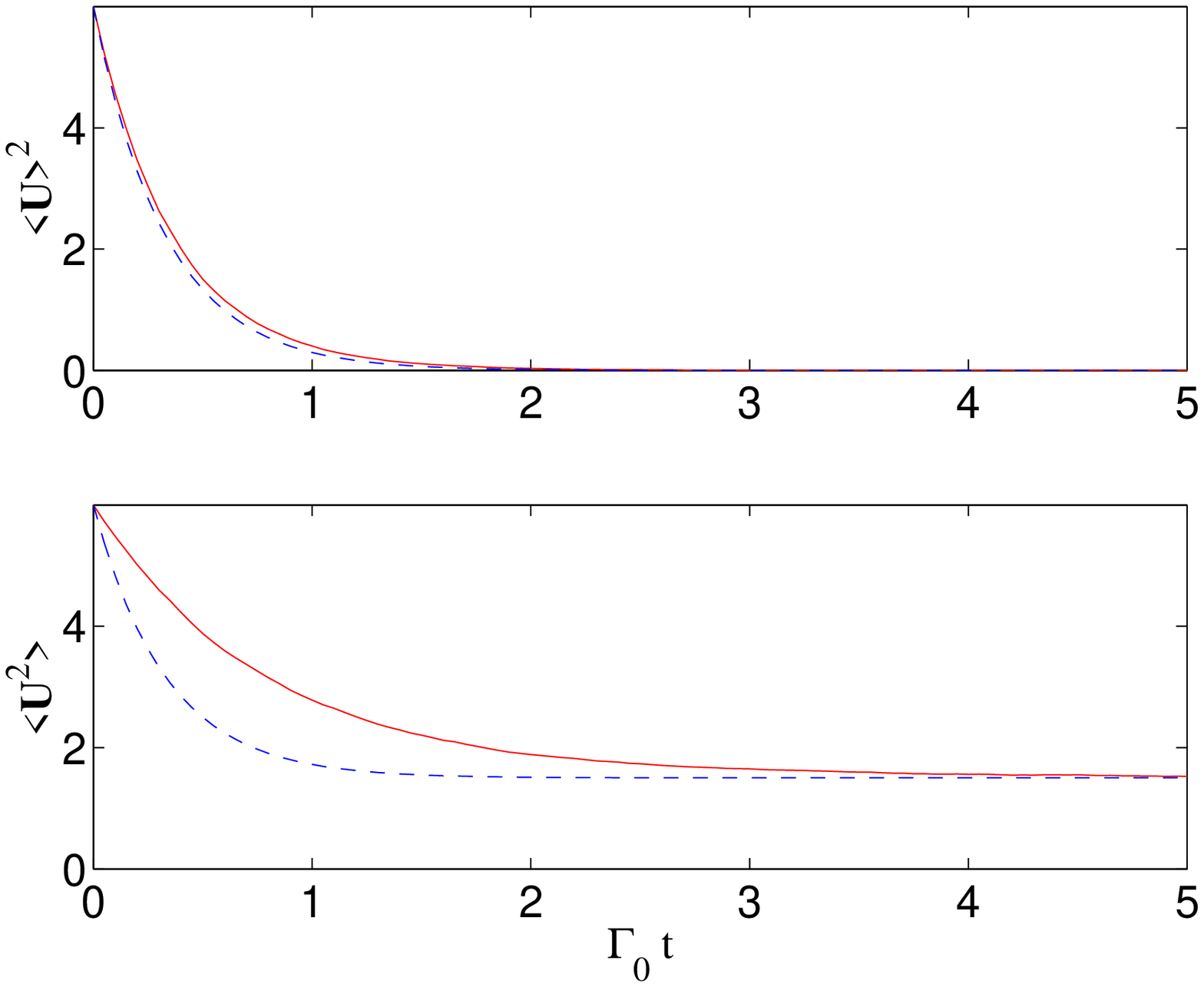}
\caption{(Color online) The same as Fig.~\ref{ensemble-1} for
$m/M=1$. \label{ensemble-2}}
\end{center}
\end{figure}

\begin{figure}[htb]
\begin{center}
\includegraphics[width=0.8\linewidth]{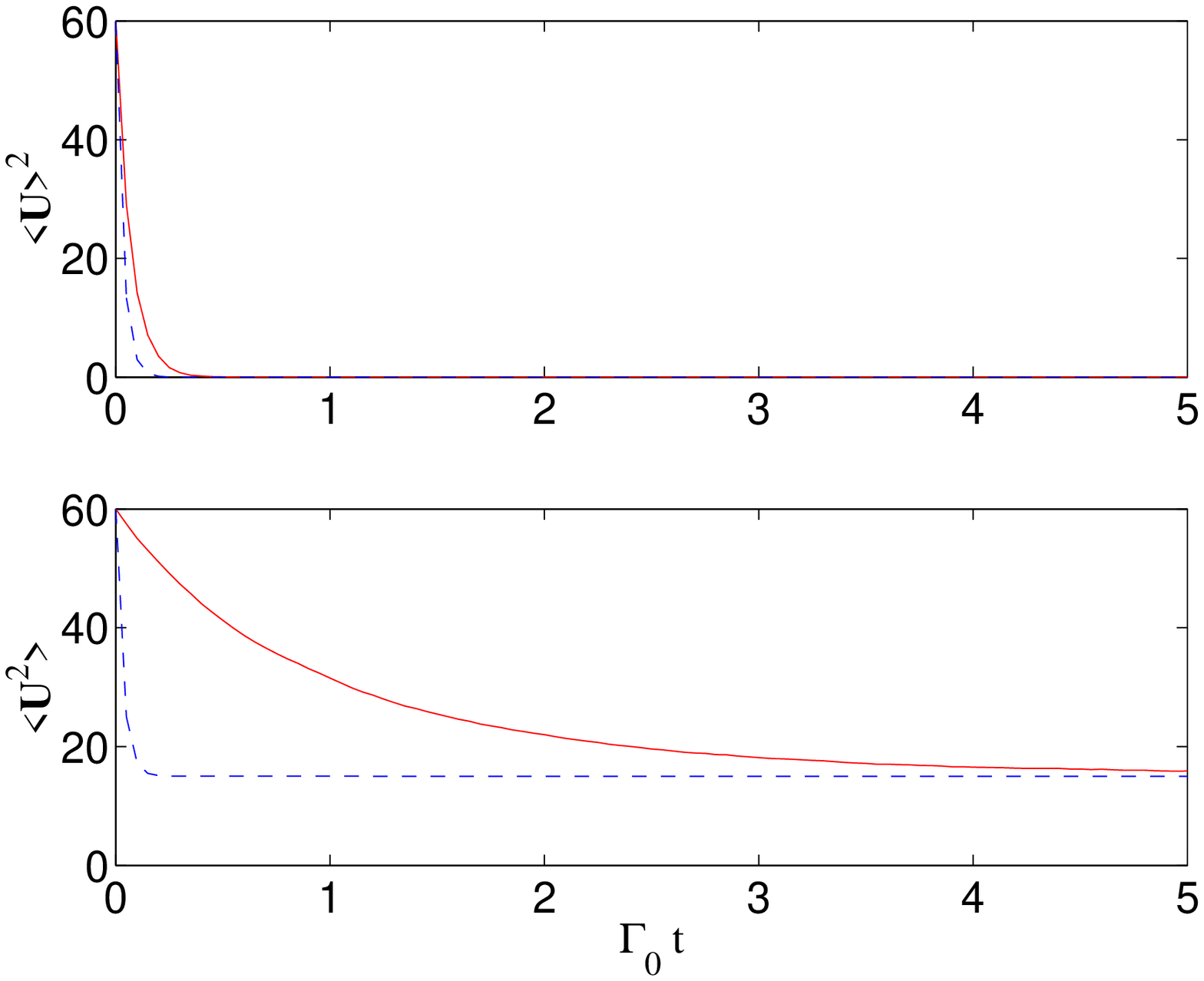}
\caption{(Color online) The same as Fig.~\ref{ensemble-1} for
$m/M=10$. \label{ensemble-3}}
\end{center}
\end{figure}

Figure \ref{fig-real} shows a single realization of the process
${\mathbf{U}}(t)$ as a 3D plot. One observes that already a few
momentum kicks drive the test particle into the vicinity of the
equilibrium value. Statistical estimates for the quantities
$\langle{\mathbf{U}}(t)\rangle^2$ and
$\langle{\mathbf{U}}(t)^2\rangle$ for various values of $m/M$ are
shown in Figs.~\ref{ensemble-1}, \ref{ensemble-2} and
\ref{ensemble-3}. In these simulations we have taken a sharp
initial state proportional to $(1,0,0)$. We see a nice relaxation
to the respective equilibrium values
$\langle{\mathbf{U}}\rangle^2_{\mathrm{eq}}=0$ and
$\langle{\mathbf{U}}^2\rangle_{\mathrm{eq}}=3m/2M$ for all
parameter combinations used.

A simple approximation of the relaxation dynamics can be obtained
in the limiting case $m/M \ll 1$. According to
Ref.~\cite{Vacchini2007d} one then finds
\begin{equation} \label{APPROX-1}
 \langle{\mathbf{U}}(t)\rangle^2 \approx \langle{\mathbf{U}}(0)\rangle^2
 e^{-\gamma_R t}
\end{equation}
and
\begin{equation} \label{APPROX-2}
 \langle{\mathbf{U}}^2(t)\rangle \approx
 \left[\langle{\mathbf{U}}^2(0)\rangle - \langle{\mathbf{U}}^2\rangle_{\mathrm{eq}} \right]
 e^{-\gamma_R t} + \langle{\mathbf{U}}^2\rangle_{\mathrm{eq}},
\end{equation}
where the relaxation rate is given by
\begin{equation} \label{LAMBDA}
 \gamma_R = \frac{16}{3\sqrt{\pi}} \frac{m}{M} \Gamma_0.
\end{equation}
We see from the figures that the
dynamics of the mean squared momentum $\langle{\mathbf{U}}^2(t)\rangle$ strongly
deviates from these approximations if $m/M$ is not small, while the
behavior of the squared
mean momentum $\langle{\mathbf{U}}(t)\rangle^2$ is still well
approximated by Eq.\eqref{LAMBDA}.
A
further characteristic feature is that for $m/M \gg 1$ the squared
mean momentum $\langle{\mathbf{U}}(t)\rangle^2$ decays much faster
than the mean squared momentum $\langle{\mathbf{U}}^2(t)\rangle$.
This means that the relaxation of the momentum and of the kinetic
energy of the particle are characterized by different decay times,
by contrast to typical master equations for quantum Brownian
motion.

\subsubsection{Gaussian scattering cross section}
\begin{figure}[htb]
\begin{center}
\includegraphics[width=0.8\linewidth]{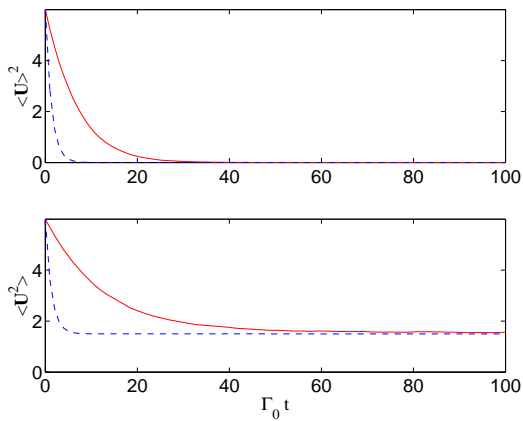}
\caption{(Color online) Averages over $10^4$ realizations for a
Gaussian cross section with $m/M=1$ and $a=1$. The broken lines
represent the approximate relaxation dynamics with the rate given
by Eq.~(\ref{LAMBDA-GAUSS}). \label{gauss}}
\end{center}
\end{figure}

For a Gaussian scattering cross section of the form
$\sigma(K)=\sigma\exp[-aK^2/4]$ the total transition rate was
given in Eq.~(\ref{GAMMA-GAUSS}). The corresponding joint
probability density becomes
\begin{equation} \label{R-K-XI-GAUSS}
 R(K,\xi) = \frac{\Gamma_0}{2\sqrt{\pi}\Gamma(U)}
 K \exp\left[ -\left(\frac{K}{2}+U\xi\right)^2 -\frac{a}{4}K^2
 \right].
\end{equation}
The qualitative features of this density are the same as for the
case of a constant cross section. Simulation results for $m/M=1$
and $a=1$ are shown in Fig.~\ref{gauss}. As expected we see the
relaxation to the equilibrium values predicted by the stationary
solution. In the case $m/M \ll 1$ the approximations given by
Eqs.~(\ref{APPROX-1}) and (\ref{APPROX-2}) are again valid, where
the relaxation rate now takes the form
\begin{equation} \label{LAMBDA-GAUSS}
 \gamma_R = \frac{16}{3\sqrt{\pi}} \frac{m}{M} \frac{1}{(1+a)^2}
 \Gamma_0.
\end{equation}

\subsubsection{Cumulants of higher order}
The stochastic process ${\mathbf{U}}(t)$ describing the particle
momentum is nonlinear in the sense that large non-Gaussian
fluctuations are dynamically generated. Starting from a Gaussian
initial state we end up for long times with a Gaussian final
(equilibrium) state. However, for intermediate times one observes
strong deviations from Gaussian statistics.

\begin{figure}[htb]
\begin{center}
\includegraphics[width=0.8\linewidth]{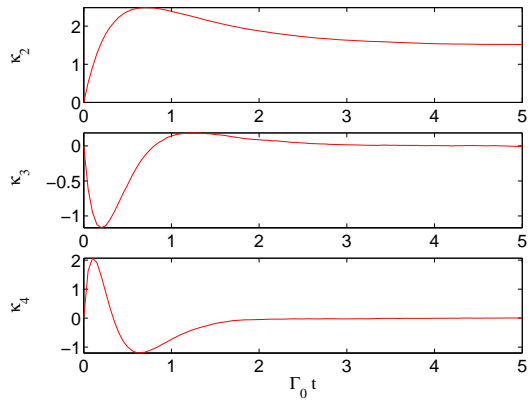}
\caption{(Color online) Cumulants of second, third and fourth
order obtained from averages over $10^5$ realizations for a
constant cross section and $m/M=1$. \label{fig-cumulants}}
\end{center}
\end{figure}

To investigate these deviations one has to consider higher moments
of the components $U_i(t)$, $i=1,2,3$, of the test particle
momentum. To this aim we have determined the time-dependence of
the cumulants of second, third and fourth order (summed over the
components of the momentum),
\begin{eqnarray}
 \kappa_2 &=& \sum_i \left\langle ( U_i-\langle U_i \rangle )^2 \right\rangle, \\
 \kappa_3 &=& \sum_i \left\langle ( U_i-\langle U_i \rangle )^3 \right\rangle, \\
 \kappa_4 &=& \sum_i \left[
 \left\langle ( U_i-\langle U_i \rangle )^4 \right\rangle
 -3 \left\langle ( U_i-\langle U_i \rangle )^2 \right\rangle^2 \right].
\end{eqnarray}
For a Gaussian distribution all cumulants of order larger than two
vanish identically. Simulation results are shown in
Fig.~\ref{fig-cumulants}. We indeed see the emergence of large
non-Gaussian fluctuations for which the cumulants $\kappa_3$ and
$\kappa_4$ are of the same order of magnitude as the variance
$\kappa_2$.

\subsection{Decoherence effects}\label{sec:decoherence}
As mentioned in Sec.~\ref{QME} diagonal matrix elements in the
momentum representation of the quantum linear Boltzmann equation
coincide with the classical expression. Typical quantum features
are therefore linked to the off-diagonal matrix elements and their
behavior in time. The suppression of these matrix elements
corresponds to a transition from the quantum to the classical
regime. It is therefore of interest to consider the evolution in
time of a superposition state. As argued in Sec.~\ref{MONTE-CARLO}
superpositions of momentum eigenstates are preserved in the course
of the time evolution. This implies that an initial state of the
form
\begin{equation} \label{eq:18}
 |\psi (0)\rangle = \sum_{i=1}^{N} \alpha_i (0) |\mathbf{U}_{i} (0)\rangle,
\end{equation}
where the amplitudes satisfy the normalization condition
$\sum_{i=1}^{N} |\alpha_i (0)|^2=1$, can be expressed at time $t$ as
\begin{equation}
   \label{eq:19}
   |\psi (t)\rangle =\sum_{i=1}^{N} \alpha_i (t) |\mathbf{U}_{i} (t)\rangle.
\end{equation}
The stochastic state vector is therefore uniquely fixed by $N$
momenta $\{\mathbf{U}_{i} (t)\}_{i=1\ldots N}$ and $N$ complex
amplitudes $\{ \alpha_{i} (t)\}_{i=1\ldots N}$ obeying the
normalization condition $\sum_{i=1}^{N} |\alpha_i (t)|^2=1$ for
any $t$.

\subsubsection{Simulation method}
Suppose that at time $t_0$ a jump into the state
\begin{equation}
   \label{eq:20}
   |\psi (t_0)\rangle =\sum_{i=1}^{N} \alpha_i (t_0) |\mathbf{U}_{i} (t_0)\rangle
\end{equation}
occurred. The deterministic time evolution before the next jump
generated by the effective Hamiltonian~\eqref{eff} is described by
\begin{equation} \label{eq:21}
 |\psi(t_0+\tau)\rangle =\sum_{i=1}^{N} \alpha_i (t_0+\tau) |\mathbf{U}_{i}(t_0)\rangle.
\end{equation}
Hence, the momenta ${\mathbf{U}}_i$ stay constant while the
dynamics of the amplitudes is given by
\begin{equation} \label{eq:22}
 \alpha_i(t_0+\tau) = \frac{e^{-i E(U_i(t_0))\tau/\hbar}
   e^{-\Gamma(U_i(t_0))\tau/2}}{\sqrt{\sum_j
   |\alpha_j(t_0)|^2e^{-\Gamma(U_j(t_0))\tau}}} \alpha_i (t_0),
\end{equation}
where $E(U) = P^2/2M = Mv_{\mathrm{mp}}^2U^2/2$. The waiting time
distribution $F(\tau)$ for the next jump is now represented by a
sum of exponential functions,
\begin{equation} \label{eq:23}
 F(\tau) = 1 - \sum_{i=1}^{N} |\alpha_i(t_0)|^2e^{-\Gamma(U_i(t_0))\tau}.
\end{equation}
The jump at time $t=t_0+\tau$ is described by the replacements
\begin{eqnarray}
 {\mathbf{U}}_i(t_0) &\longrightarrow& {\mathbf{U}}_i(t_0) + \frac{m_*}{M}{\mathbf{K}}, \\
 \alpha_i(t_0+\tau) &\longrightarrow& f_i \alpha_i(t_0+\tau),
 \label{ALPHA-TRANS}
\end{eqnarray}
where the factors $f_i$ are given by
\begin{equation} \label{F-I}
 f_i =
 \frac{e^{-\frac{1}{2}\left(K/2+{\mathbf{K}}\cdot{\mathbf{U}_i(t_0)}/K\right)^2}}
 {\sqrt{\sum_j |\alpha_j(t_0+\tau)|^2 e^{-\left(K/2+{\mathbf{K}}\cdot{\mathbf{U}_j(t_0)}/K\right)^2}}}.
\end{equation}
The momentum transfer ${\mathbf{K}}$ in these formulas is to be
drawn from the corresponding probability density
$R({\mathbf{K}})$. To determine ${\mathbf{K}}$ one proceeds as
follows. First, one draws an index $i\in\{1,2,\ldots,N\}$ with
probability
\begin{equation}
 p_i = \frac{|\alpha_i(t_0)|^2e^{-\Gamma(U_i(t_0))\tau}\Gamma(U_i(t_0))}
 {\sum_j |\alpha_j(t_0)|^2e^{-\Gamma(U_j(t_0))\tau}\Gamma(U_j(t_0))}.
\end{equation}
Given $i$ one then draws a momentum transfer ${\mathbf{K}}$ that
follows the probability density $R(K,\xi_i)$ [see
Eq.~(\ref{R-K-XI})], where now $\xi_i$ represents the cosine of
the angle between ${\mathbf{K}}$ and ${\mathbf{U}}_i(t_0)$.

\subsubsection{Decay of coherences}
\begin{figure}[htb]
\begin{center}
\includegraphics[width=0.8\linewidth]{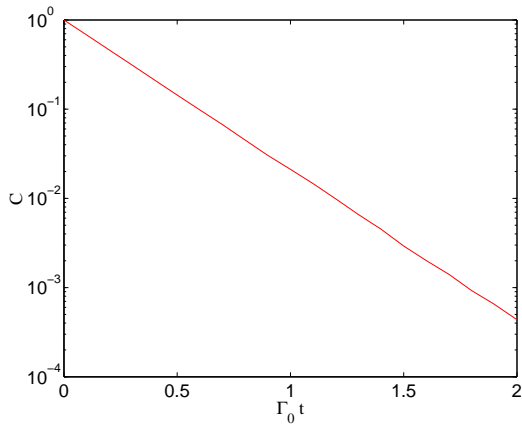}
\caption{(Color online) Semilogarithmic plot of the coherence
$C(t)$ obtained from an average over $10^5$ realizations for
$m/M=1$ (constant cross section). \label{fig-Ct}}
\end{center}
\end{figure}

We have used the simulation algorithm described above to study the
loss of coherence of  an initial state of the form
\begin{equation} \label{INIT-DEC}
 |\psi(0)\rangle = \alpha_1(0)|{\mathbf{U}}_1(0)\rangle
 + \alpha_2(0)|{\mathbf{U}}_2(0)\rangle,
\end{equation}
given by a superposition of two momentum eigenvectors. The initial
state is to be drawn from a given initial distribution for the
momenta ${\mathbf{U}}_{1,2}(0)$ and amplitudes $\alpha_{1,2}(0)$.
In the simulations shown below we have taken sharp opposite
initial momenta,
\begin{equation} \label{INIT-1}
 {\mathbf{U}}_1(0) = -{\mathbf{U}}_2(0) \equiv {\mathbf{U}}_0,
\end{equation}
and equal amplitudes,
\begin{equation} \label{INIT-2}
 \alpha_1(0) = \alpha_2(0) = \frac{1}{\sqrt{2}}.
\end{equation}
Such an initial state is a balanced coherent superposition of two
momentum eigenvectors separated by twice $U_0 \equiv \Delta
P/p_{\mathrm{mp}}$, where $p_{\mathrm{mp}}=m v_{\mathrm{mp}}$ is
the most probable momentum of the gas particle at temperature
$T=1/k_{\text{B}}\beta$. In order to study quantitatively the loss
of coherence we have used the simulation algorithm to estimate the
expectation value
\begin{equation}
 C(t) = {\mathbb{E}}\left[\frac{|\alpha_1(t)\alpha_2^*(t)|}{|\alpha_1(0)\alpha_2^*(0)|}\right]
 = 2{\mathbb{E}}\left[|\alpha_1(t)\alpha_2^*(t)|\right].
\end{equation}
This quantity represents the average of the absolute values of the
coefficients in front of the off-diagonal matrix elements of the
test particle's statistical operator in the momentum
representation, divided by its initial value. Hence, $C(t)$
provides a measure for the degree of the coherence of the state of
the test particle. An example for the dynamical behavior of $C(t)$
is shown in Fig.~\ref{fig-Ct}. In this figure we have used a sharp
initial state given by ${\mathbf{U}}_0=(0,0,4)$. We clearly see an
exponential decay of the coherence $C(t)$ over several orders of
magnitude.

\begin{figure}[htb]
\begin{center}
\includegraphics[width=0.8\linewidth]{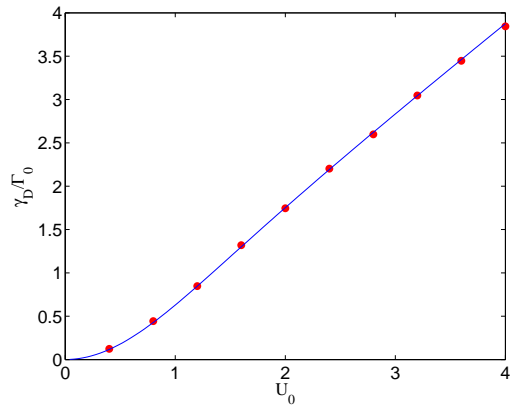}
\caption{(Color online) The decoherence rate $\gamma_D$ in units
of $\Gamma_0$ as a function of the initial momentum $U_0$. Points:
Least squares fits of the simulation data for $m/M=1$. Continuous
line: Analytical estimate given by Eq.~\eqref{DECRATE-CONSTANT}.
 \label{fig-decrate}}
\end{center}
\end{figure}

Assuming that an exponential decay of the coherence holds true, we
define the decoherence rate $\gamma_D$ by means of
\begin{equation} \label{DEF-DEC-RATE}
 C(t) = e^{-\gamma_D t}.
\end{equation}
An analytical approximation for $\gamma_D$ can be found with the
help of the following argument. We take a sufficiently small time
$t$ such that we can approximate
\begin{equation} \label{DERIV-1}
 C(t) \approx 1 - \gamma_D t
 \approx 1-\Gamma(U_0)t + \Gamma(U_0)t \langle f_1 f_2 \rangle.
\end{equation}
Here, $\Gamma(U_1(0))=\Gamma(U_2(0))=\Gamma(U_0)$ represents the
total rate for a transition out of the given initial state
(\ref{INIT-DEC}). Hence, $\Gamma(U_0)t$ represents the probability
for a jump within time $t$, while $1-\Gamma(U_0)t$ is the
probability that no jump occurs. Using Eqs.~(\ref{INIT-1}) and
(\ref{INIT-2}) we see from Eq.~(\ref{eq:22}) that $C$ does not
change during the deterministic drift. On the other hand, if a
jump does occur then $C$ changes from its initial value $C=1$ to
$C=f_1f_2$ as may be seen from Eq.~(\ref{ALPHA-TRANS}). According
to Eq.~(\ref{F-I}) we have
\begin{equation}
 f_1f_2 = \frac{2}{e^{{\mathbf{K}}\cdot{\mathbf{U}}_0}+e^{-{\mathbf{K}}\cdot{\mathbf{U}}_0}}.
\end{equation}
Thus, Eq.~(\ref{DERIV-1}) represents the change of $C$ as a result
of two alternatives, namely that a jump does occur or that it does
not (for small enough $t$ we can have at most one jump). Finally,
$\langle f_1f_2 \rangle$ denotes the average of $f_1f_2$ taken
over the possible momentum transfers during the first jump. We
therefore get
\begin{eqnarray} \label{DERIV-2}
 \gamma_D &\approx& \Gamma(U_0) ( 1 - \langle f_1f_2 \rangle ) \nonumber \\
 &=& \Gamma(U_0) \left\langle
 1- \mathrm{Sech}({{\mathbf{K}}\cdot{\mathbf{U}}_0})
 \right\rangle.
\end{eqnarray}
This formula can be analytically evaluated in several cases. For a
constant scattering cross section one finds exploiting
Eq.~(\ref{R-K-XI}),
\begin{equation} \label{DECRATE-CONSTANT}
 \gamma_D = \Gamma (U_0) -\Gamma_0\frac{\mathrm{erf} (U_0)}{U_0}.
\end{equation}
Fig.~\ref{fig-decrate} demonstrates the extremely good agreement
between the stochastic simulation results and this analytical
approximation. The Monte Carlo estimates for the decoherence rate
have been obtained by least squares fits to the simulation data
within the time interval in which the coherence decays to $1\%$ of
its initial value. For all cases shown we find a nearly perfect
exponential decay. Our argument shows that it is just the real
scattering events that mainly cause the decoherence during the
early phase of the dynamics, by contrast to the virtual
transitions described by the non-Hermitian drift Hamiltonian.

\begin{figure}[htb]
\begin{center}
\includegraphics[width=0.8\linewidth]{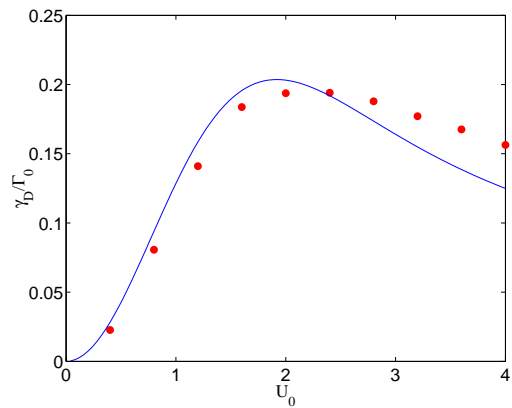}
\caption{(Color online) The same as Fig.~\ref{fig-decrate} for a
Gaussian cross section with $a=1$. The continuous line represents
the analytical estimate given by Eq.~\eqref{DECRATE-GAUSS}.
 \label{fig-decrate-gauss}}
\end{center}
\end{figure}

For a Gaussian cross section Eq.~(\ref{DERIV-2}) leads to
\begin{equation} \label{DECRATE-GAUSS}
 \gamma_D = \Gamma (U_0) -\Gamma_0\frac{\mathrm{erf} (U_0)}{U_0}
 \frac{1}{1+a}.
\end{equation}
In Fig.~\ref{fig-decrate-gauss} we compare this expression with
the simulation results. Although the quantitative agreement is
obviously not as good as in the case of a constant cross section,
the analytical formula (\ref{DECRATE-GAUSS}) still yields a
reasonable estimate for the order of magnitude of the decoherence
rate.

The above results enable us to study the relationship between the
time scales characterizing the different physical phenomena. For
the case of $m/M\ll 1$ the time scale for relaxation is given by
Eq.~(\ref{LAMBDA}). Considering a wide separation in momentum of
the initial superposition on the scale set by the momentum of the
gas particles, so that $U_0=\Delta P/p_{\mathrm{mp}}\gg 1$, one
can consider the corresponding limiting value $\Gamma_0 U_0$ of
Eq.~\eqref{DECRATE-CONSTANT}, and the ratio of the decoherence
rate to the relaxation rate is given by
\begin{equation}
 \frac{\gamma_D}{\gamma_R} \approx \frac{3\sqrt{\pi}}{16}
 \frac{M}{m} U_0 \gg 1.
\end{equation}
This shows that the decoherence time $\gamma_D^{-1}$ for the loss
of coherence in momentum space can be much smaller than the
relaxation time $\gamma_R^{-1}$, i.~e. than the time it takes for
the relaxation of the energy of the test particle.

It is also of interest to compare the relaxation timescale with
the decoherence rate $\eta_D$ of a superposition of position
eigenstates, widely separated on the scale set by the thermal
wavelength of the gas particles, which corresponds to $\hbar$
divided by the typical momentum transfer, so that $\Delta
X/\lambda_{\mathrm{th}}\gg 1$. In such a case for $m/M\ll 1$ the
decoherence rate is set  by the total transition rate
\cite{Joos1985a,Gallis1990a,Hornberger2003a,Hornberger2003b,Vacchini2004a,Vacchini2005b},
so that for a test particle much slower than the gas particles one
has $\eta_D\approx 2\Gamma_0/\sqrt{\pi}$. Thus, the ratio
\begin{equation}
 \frac{\eta_D}{\gamma_R} \approx \frac{3}{8}
 \frac{M}{m} \gg 1
\end{equation}
shows again that the decoherence time $\eta_D^{-1}$ for the loss
of coherence in position space can be much smaller than the
relaxation time $\gamma_R^{-1}$, but still longer than the
decoherence time in momentum space  $\gamma_D^{-1}$.

\section{Conclusions}\label{sec:conclusions}

We have developed a stochastic unravelling of the quantum linear
Boltzmann equation which leads to an efficient Monte Carlo
simulation technique, despite the appearance in the equation
itself of quite complicated operator-valued expressions. The
latter are responsible for deviations from Gaussian statistics, at
variance with typical master equations used for the description of
quantum Brownian motion. A crucial feature of the method is that
the developed algorithms fully exploit the translation covariance
and thus allow full three-dimensional stochastic simulations of
the quantum Boltzmann equation. In particular, the method does not
require the introduction of a discretization in momentum space, so
that the continuous sum over the Lindblad operators in
Eq.~\eqref{QMEQ} can be exactly accounted for.

The method proposed here suggests many further physically relevant
applications. For example, one can extend the algorithm to the
regime where effects from the Bose or Fermi statistics of the
quantum gas come into play. In fact, starting from
Eqs.~\eqref{eq:2} and \eqref{eq:1} one can analytically work out
the corresponding expressions of the dynamic structure factor for
a free gas of particles obeying Bose or Fermi statistics, coming
to \cite{Vacchini2001b}
\begin{eqnarray*} \label{B-F}
 S_{\rm \scriptscriptstyle B/F}(Q,E) &=&
 \frac{1}{(2\pi\hbar)^3} \frac{2\pi m^2}{n_{\mathrm{gas}}\beta Q}
 \frac{\mp 1}{1-e^{\beta E}} \\
 && \times \ln \left[
 \frac{1 \mp z\exp\left[-\frac{\beta}{8m}\frac{(2mE+Q^2)^2}{Q^2}\right]}
 {1 \mp z\exp\left[-\frac{\beta}{8m}\frac{(2mE-Q^2)^2}{Q^2}\right]}
 \right], \nonumber
\end{eqnarray*}
where the upper signs refer to the Bose case and the lower signs
to the Fermi case, and $z$ denotes the fugacity of the gas. By use
of this expression the algorithm thus enables Monte Carlo
simulations of the behavior of test particles in a Bose or Fermi
gas to identify genuine effects of the quantum statistics. More
generally, the method may also be applied to an interacting
quantum gas provided an (at least approximate) expression for the
dynamic structure factor $S({\mathbf{Q}},{\mathbf{P}})$ is known.

A further example is the investigation of the important problem of
decoherence in position space. This can be done by use of initial
states representing superpositions of localized wave packets, with
the aim of determining the corresponding position space
decoherence time scales. Localized wave packets may of course be
represented by introducing a discretization of position space.
However, it seems that it is much more efficient to invoke the
translation covariance and to describe spatially localized states
by appropriate superpositions of momentum eigenstates as discussed
in Sec.~\ref{sec:decoherence}, or, more generally, by
superpositions of wave packets localized in momentum space.
Further examples of application include the extension of the
method to the determination of multitime correlation functions, to
the treatment of particles with internal degrees of freedom, and
to the case of an operator-valued scattering amplitude, i.~e. to
the case that the scattering cross section depends on the momentum
of the incoming test particle.

\begin{acknowledgments}
Bassano Vacchini would like to thank Klaus Hornberger and Ludovico
Lanz for many fruitful discussions. The work was partially
supported by the Italian MIUR under PRIN05.
\end{acknowledgments}


\end{document}